\documentclass[twocolumn]{aastex62}

\usepackage{epstopdf}
\usepackage{rotating}
\usepackage{natbib}
\usepackage{amsmath}

\newcommand{\upp}

\begin{document}

\title{The Number of Possible CETIs within Our Galaxy and the Communication Probability among These CETIs}

\correspondingauthor{He Gao}
\email{gaohe@bnu.edu.cn}

\author{Wenjie Song}
\affil{Department of Astronomy ,
Beijing Normal University, Beijing, People's Republic of China;}

\author[0000-0002-3100-6558]{He Gao}
\affiliation{Department of Astronomy ,
Beijing Normal University, Beijing, People's Republic of China;}

\begin{abstract}
As the only known intelligent civilization, human beings are always curious about the existence of other communicating extraterrestrial intelligent civilizations (CETIs). Based on the latest astrophysical information, we carry out Monte Carlo simulations to estimate the number of possible CETIs within our Galaxy and the communication probability among them. Two poorly known parameters have a great impact on the results. One is the probability of life appearing on terrestrial planets and eventually evolving a into CETI ($f_c$), and the other determines at what stage of their host star's evolution CETIs would be born ($F$). In order to ensure the completeness of the simulation, we consider a variety of combinations of $f_c$ and $F$. Our results indicate that for optimistic situations (e.g. $F=25\%$ and $f_c=0.1\%$), there could be $42777_{-369}^{+267}$ CETIs and they need to survive for $3_{-2}^{+17}$ yr ($2000_{-1400}^{+2000}$ yr) to achieve one-way communication (two-way communication). In this case, human beings need to survive $0.3_{-0.298}^{+0.6}$ Myr to receive one alien signal. For pessimistic situations (e.g. $F=75\%$ and $f_c=0.001\%$), only $111_{-17}^{+28}$ CETIs could be born and they need to survive for $0.8_{-0.796}^{+1.2}$ Myr ($0.9_{-0.88}^{+4.1}$ Myr) to achieve one-way communication (two-way communication). In this case, human beings need to survive $50_{-49.6}^{+250}$ Myr to receive one signal from other CETIs. Our results may quantitatively explain why we have not detected any alien signals so far. The uncertainty of the results has been discussed in detail and would be alleviated with the further improvement of our astronomical observation ability in the future. 

\end{abstract}


\section{Introduction} \label{sec:intro}
As the only advanced intelligent civilization on the Earth, one of the most puzzling questions for humans is whether our existence is unique. There have been many studies on extraterrestrial civilization in the past few decades. The modern era of the search for extraterrestrial intelligence (SETI) began in 1960 when \citet{1961PhT....14d..40D} began the first SETI program in radio band, Project Ozma. Besides, Frank Drake later proposed a formula to estimate the number of communicating civilizations in the Galaxy, named the Drank Equation \citep{1965Frank}. Following in-depth study, some people think that maybe we can not only listen but also actively send signals. In 1971, the first Soviet-American Conference on Communication with Extraterrestrial Intelligence (CETI) was held at the Byurakan Astrophysical Observatory of the Armenian Academy of Sciences, USSR\citep{1972Byurakan}. The best-known CETI experiment was the 1974 Arecibo message composed by Drake and Sagan \citep{Carl}. However, recently some changes have taken place in the meaning of CETI. In recent literature (and henceforth in this work), CETI refers to a communicating extraterrestrial intelligent civilization. In addition, it is always controversial whether people should passively search for technical signatures or seek responses. In order to distinguish these search strategies, they are called passive SETI (usually just called SETI) and active SETI (which is usually called messaging extraterrestrial intelligence (METI); \cite{2006physics..10031Z}). \citet{2021AcAau.188..203W} reviewed some of the various observational search strategies for SETI.
\par We have always wanted to know the answers to the following questions. First, how many CETIs exist in the Milky Way? This is a challenging problem. We can only learn from a single known data point (ourselves) without knowing the sample mean or standard deviation. The process of this attempted extrapolation from $N=1$ would seem to push the integrity of logic to its limits \citep{2020ApJ...896...58W}. However, we can propose models based on reasonable logical assumptions, which may produce plausible estimates of the occurrence rate of communicating civilizations. Most studies on this problem are based on the Drake equation. The obvious difficulty of this method is that it is uncertain and unpredictable to quantify the probability that life may appear on a suitable planet and eventually develop into an advanced communicating civilization. Due to the development of astrophysics and our knowledge of star formation and planetary systems, especially the discovery of more and more exoplanets and the new data on star formation rate (SFR), it is possible to conduct a quantitative study on this problem. 

In principle, there are two approaches to deal with the variables in the Drake equation: (1) applying proper distributions as a prior and providing a probabilistic solution by using Bayesian method (see \cite{Maccone2010,Maccone2011,2019Bloetscher} for detailed descriptions); (2) based on the Copernicus principle, taking the latest astrophysical observation results as the distribution function of the corresponding variables; for instance, using the established SFR data of distant galaxies to manifest the SFR history of the Milky Way (see \cite{2020ApJ...896...58W} for a detailed description). 

For the first approach, \citet{2019Bloetscher} provides a probabilistic solution for the Drake equation by using predictive Bayesian Markov Chain Monte Carlo methods and finds
the maximum number of contemporary civilizations might be as few as a 1000. For the other approach, \citet{2020ApJ...896...58W} investigate the possible number of CETIs in the Milky Way under different conditions by using new data about the history of star formation in the Milky Way. They propose two scenarios; one is that intelligent life can only form on an Earth-like planet in a habitable zone after the star is at least 5 Gyr old (which they call the Weak Astrobiological Copernican scenario), and the other is that life must form between 4.5 and 5.5 Gyr (the Strong Astrobiological Copernican scenario). Their results indicate that there should be at least $36^{+175}_{-32}$ civilizations in the Galaxy.

\par If their results are reliable, the second question we are concerned with is the communication probability among these CETIs. In particular, the question of whether we humans can receive signals from other CETIs is important. Since the appearance of CETIs is an accidental event with great randomness, it is not easy to solve these problems by using analytic methods. For the purpose of this work, we suggest using the latest astrophysical information (i.e. the second approach) and Monte Carlo simulation to investigate the possible number of CETIs and the communication probability among CETIs. The advantage of this method is that we can simulate for many times so as to better estimate the uncertainty of the results. In the simulation, we trace the star formation history of the Milky Way to generate proper numbers of stars at specific times. Based on the knowledge of the initial mass function (IMF), metallicity distribution, and the relative fraction of stars within each region of Milky Way ($r=15 \rm{kpc}$, $z=4\rm{kpc}$), we assign the mass, metallicity, and locations to each star. For each star, the probability of forming CETIs would be $f_{\rm{CETI}}=f_{\rm{p}}\cdot f_{\rm{c}}$, where $f_{\rm{p}}$ is the average probability of terrestrial planets appearing in the habitable zones of stars, and $f_{\rm{c}}$ is the probability of life appearing on terrestrial planets and eventually evolving into a CETI. For stars that can give birth to CETI, we assign the birth time for each CETI as $T_{\rm{c}}=T_{\rm{s}}+F\cdot \tau_i$, where $T_{\rm{s}}$ is the birth time of the host star, $\tau_i$ is the main-sequence lifetime of the host star, and $F$ decides at what stage of star evolution the CETI was born. Finally we assign the communication lifetime ($\Delta T$) to each CETI, and then estimate the communication probability among these CETIs. Obviously, the communication lifetime of CETIs plays an important role. If the communication lifetime of CETIs is long enough, the probability could approach unity. This work focuses on investigating the critical ${\Delta T}$ value required for any CETI successfully receiving signals from other CETIs, in particular, quantitatively predicting how long we humans need to survive to discover CETIs. We will systematically describe our simulation method in Section \ref{2}. Our results and discussions are shown in Section \ref{3}-\ref{4}.

\section{The Simulation Method}
\label{2}
\subsection{Star Formation History and the Distribution of Stars}
First, we need to simulate the formation history of stars. Because the star formation history of the local galaxy group is quite consistent with the global history \citep[e.g.][]{2014ApJ...789..147W}, as proposed by \citet{2020ApJ...896...58W}, the established SFR data of distant galaxies \citep[e.g.][]{2014ARA&A..52..415M} can be used to manifest the SFR history of the Milky Way. The analytic form of SFR with respect to redshift has been proposed as \citep{2003MNRAS.341.1253H}
\begin{equation}
    \dot{\rho}_{*}(z)= \dot{\rho}_{*}(0)\frac{\chi^{n_1}}{1+\alpha(\chi-1)^{n_2}e^{\beta\chi^{n_3}}},
\end{equation}
where $\dot{\rho}_*(z)$ is SFR at redshift $z$ (in units of $M_{\odot} \rm ~Mpc^{-3}~yr^{-1}$), $\dot{\rho}_*(0)$ is the present-day SFR, and $\chi(z)$ is defined as
\begin{equation}
    \chi(z)=\left[\frac{H(z)}{H_0}\right]^{2/3},\notag
\end{equation}
where $H(z)=H_0[\Omega_M(1+z)^3+(1-\Omega_M-\Omega_\Lambda)(1+z)^2+\Omega_\Lambda]^{\frac{1}{2}}$ is the Hubble parameter. Throughout this paper, we assume the Planck cosmology \citep{planck15} as a fiducial model, with ${\Omega _{\rm{m}}} = 0.307$, ${\Omega _\Lambda } = 0.693$, and ${H_0} = 67$ ${\rm{km}}$ ${{\rm{s}}^{ - 1}}$ ${\rm{Mp}}{{\rm{c}}^{ - 1}}$.

For other parameters, we adopt the best-fitting results given by \citet{2020ApJ...896...58W}, namely $\rm{\alpha}=0.528$, $\rm{\beta}=2.36$, $\dot{\rho}_*(0)=0.330$, $n_1=5.91$, $n_2=-0.3508$, and $n_3=1.22$. Converting redshift into lookback time ($t_{\rm lb}(z)$), we can then simulate the total mass of stars produced in any time period ($t_{\rm lb}(z)$, $t_{\rm lb}(z)+\Delta t$) as 
\begin{equation}
    M_{\rm formed}=\dot{\rho}_*(z) \times 10^8 M_{\odot} \rm Mpc^{-3},
\end{equation}
where the simulation time-step $\Delta t$ is adopted as $10^8$ yrs. By introducing the IMF, which is a function that describes the mass distribution of stars in a star cluster or in a galaxy, we can know the total number of stars formed in any time period ($t_{\rm lb}(z)$, $t_{\rm lb}(z)+\Delta t$). Here we choose the Salpeter IMF \citep{1955ApJ...121..161S},
\begin{equation}
    \frac{dN}{dM}=kM^{\alpha},~~~M_{\rm low}<M<M_{\rm upp},
\end{equation}
where $k$ is the normalization constant and \begin{math}\alpha=-2.35\end{math}. Here we adopt \begin{math} M_{\rm low}=0.08 \ M_ {\odot}\end{math} (the minimum mass required for hydrogen fusion; \citep{2006Sci...313..936R} and \begin{math}M_{\rm upp}=100\ M_{\odot}\end{math} \citep{2005Natur.434..148K,2006Sci...313..936R}. In this case, we will simulate $N_{\rm formed}$ stars with birth time $t_{\rm lb}(z)$,
where
\begin{equation}
    N_{\rm formed} = (\frac{\alpha+2}{\alpha+1})(\frac{M_{\rm upp}^{\alpha+1}-M_{\rm low}^{\alpha+1}}{M_{\rm upp}^{\alpha+2}-M_{\rm low}^{\alpha+2}})M_{\rm formed}
\end{equation}
and each star is randomly assigned a mass by using the IMF as the probability density distribution. 
\par Next, we will assign each star their locations in the Milky Way. A simple exponential model is adopted here to calculate the probability for each star appearing within different locations \citep{2020ApJ...896...58W}:
\begin{equation}
p(R)\propto e^{-\frac{\rm{R}}{\rm{h_R}}},
\end{equation}
\begin{equation}
p(z)\propto e^{-\frac{\rm{z}}{\rm{h_z}}},    
\end{equation}
where $h_{\rm{R}}=2.5\pm 0.4$ \rm{kpc} and ${h_{\rm{z}}}=220\sim450$ \rm{pc} present the typical scale length and height of the disk, and the full disk radius and thickness are taken as 15 $\rm{kpc}$ and 4 $\rm{kpc}$ (2 kpc above and below the midplane), respectively \citep{2016ARA&A..54..529B}.  

\subsection{CETI Forming Conditions and Appearance Time}
\label{2.2}
Since we have not found any other CETI yet, we can only assume that the evolution of other CETIs may have a similar path to the evolution of human beings. CETIs may be born on terrestrial planets located in a habitable zone, where the temperature around the star is not too high or too low so that liquid water on the surface of the planet can exist. Previous studies have shown a strong correlation between planet formation and host star metallicity \citep[e.g.][]{2005ApJ...622.1102F,2012Natur.486..375B}. If the metallicity is too low, the formation of planets will be inhibited. \citet{2012ApJ...751...81J} first discussed the minimum metallicity required to form a planetary system, and believed that the metallicity required to form a terrestrial planet is $Z\ge 0.1Z_\odot$. Here we adopt the results from \citet{2006MNRAS.366..899N} (see their figure 11 for details) to calculate the metallicity for stars born at different times and in different locations. In the following simulation, we focus on stars with metallicity \begin{math}[F_e/H]>-1\end{math} and believe that these stars have enough metallicity to form terrestrial planets.

\par However, it is worth noting that not all stars with enough metallicity could have terrestrial planets. Thanks to the successful operation of the Kepler mission, many insightful results have been achieved in estimating the percentage of stars of different spectral types that host planets with particular characteristics \citep{2011B,2012ApJ...745...20T,2015ApJ...807...45D}. For instance, by analyzing the FGK target star transit phenomenon with a period of less than 42 days in the Kepler database \citep{2011B}, \cite{2012ApJ...745...20T} showed that the average probability of terrestrial planets appearing in the habitable zone of FGK stars is around $\simeq (34\pm 14)\%$. \cite{2015ApJ...807...45D} analyzed data from the 4 yr of operation of the Kepler mission and suggested that the occurrence rate of terrestrial planets within the habitable zone of M stars is around $0.16^{+0.17}_{-0.07}$. Given that the probability of terrestrial planets appearing in the habitable zone of OBA stars is relatively small due to their high temperature, here we only consider that FGKM stars could host terrestrial planets in their habitable zone. For our simulated FGK stars, we randomly adopt a percentage of $34\%$ stars to be the candidates that host terrestrial planets in their habitable zone, while for our simulated M stars, we randomly adopt a percentage of $16\%$ stars to be the candidates that host terrestrial planets in their habitable zone. 

\par For each selected candidate, there will be a certain probability ($f_c$) that a CETI can be generated. Since we know nothing about how large $f_c$ should be, here we test three values of different orders of magnitude ($0.1\%$, $0.01\%$, and $0.001\%$) and analyze the influence of different values on the results. On the other hand, we assume that the birth time of CETIs is related to the main-sequence lifetime of their host stars:
\begin{equation}
    T_{\rm c}=T_{\rm s}+F\cdot \tau_i,
\end{equation}
where $T_{\rm c}$ refers to the time when a civilization has developed to the extent that it can communicate with other civilizations by sending and receiving signals, $T_{\rm{s}}$ is the birth time of the host star, $\tau_i$ is the main-sequence lifetime of the host star, and $F$ decides at what stage of star evolution the CETI was born. Here we use the relationship between main-sequence lifetime and mass to estimate $\tau_i$ as
\begin{equation}
    \begin{cases}
    \frac{\tau_i} {Gyr}{\approx}7.1(\frac{M}{M_\odot})^{-2.5}\quad\text{for }2M_{\odot}\le M \le 20M_{\odot}, \\
    \frac{\tau_i}{Gyr}{\approx}10(\frac{M}{M_\odot})^{-3} \qquad \text{for }{0.43}M_{\odot}\le M < 20M_{\odot},\\
    \frac{\tau_i}{Gyr}{\approx}43(\frac{M}{M_\odot})^{-1.3}\quad
    \text{for } M < 0.43M_{\odot}.
    \end{cases}
    \label{9}
\end{equation}
$F$ is another parameter that we know little about. Here we test three values for $F$ ($25\%$, $45.7\%$, and $75\%$) and analyze the influence of different values on the results.  Note that $F=45.7\%$ is adopted based on the value for human beings, which is the only example we have. The other two values are adopted by roughly a quarter larger or smaller than $45.7\%$. 

\subsection{Communication Probability between CETIs}

With all simulated CETIs being assigned their locations and birth times, we can estimate the communication probability among them once we assign the communication lifetime ($\Delta T$) to each CETI. Here the communication lifetime means the length of time during which civilizations could release and receive communicable signals. For simplicity, we assume $\Delta T$ is the same for all CETIs. In principle, there are two levels of communication: (1) one-way communication, that is, where one CETI receives a signal from another CETI, but does not respond to the signal; (2) two-way communication, that is, where one CETI receives the signal sent by another CETI and immediately transmits a response signal, which is also received by the other side. 

For any two CETIs, whose locations and birth times are $(X_1,Y_1,Z_1,T_1)$ and $(X_2,Y_2,Z_2,T_2)$, respectively, the conditions for them to successfully achieve one-way communication could be justified as follows:

1. Case (a): if $T_{2}>T_{1}$ and $T_{2}>T_{1}+\Delta T$, we need $T_2\le T_1+T_{\rm sp}\le T_2+\Delta T$ or $T_2\le T_1+\Delta T+T_{\rm sp}\le T_2+\Delta T$;

2. Case (b): if $T_{1}>T_{2}$ and $T_{1}>T_{2}+\Delta T$, we need $T_1\le T_2+T_{\rm sp}\le T_1+\Delta T$ or $T_1\le T_2+\Delta T+T_{\rm sp}\le T_1+\Delta T$;

3. Case (c): if $T_{2}>T_{1}$ and $T_{2}<T_{1}+\Delta T$, we need $T_2\le T_1+T_{\rm sp}\le T_2+\Delta T$ or $T_2\le T_1+\Delta T+T_{\rm sp}\le T_2+\Delta T$ or $T_1\le T_2+T_{\rm sp}\le T_1+\Delta T$;

4. Case (d): if $T_{1}>T_{2}$ and $T_{1}<T_{2}+\Delta T$, we need $T_1\le T_2+T_{\rm sp}\le T_1+\Delta T$ or $T_1\le T_2+\Delta T+T_{\rm sp}\le T_1+\Delta T$ or $T_2\le T_1+T_{\rm sp}\le T_2+\Delta T$.

Here $T_{\rm sp} ={\sqrt{(X_2-X_1)^2+(Y_2-Y_1)^2+(Z_2-Z_1)^2}}/{c}$ is the required signal-spreading time. Here we assume that the signal travels at the speed of light ($c$). 

In order to realize two-way communication, the lifetime periods for the CETI pair must intersect, and the following conditions are required:

1. Case (c'): if $T_{2}>T_{1}$ and $T_{2}<T_{1}+\Delta T$, we first need $T_2\le T_1+T_{\rm sp}\le T_2+\Delta T$ or $T_{2}\le T_1+\Delta T + T_{\rm sp}\le T_{2}+\Delta T$, then need $T_{1}\le T_{2}+T_{\rm sp} \le T_{1}+\Delta T$;

2. Case (d'): if $T_{1}>T_{2}$ and $T_{1}<T_{2}+\Delta T$, we first need $T_1\le T_2+T_{\rm sp}\le T_1+\Delta T$ or $T_1\le T_2+\Delta T+T_{\rm sp}\le T_1+\Delta T$, then need $T_2\le T_1+T_{\rm sp}\le T_2+\Delta T$.

For each CETI, we calculate whether it can achieve one-way and two-way communications with other CETIs. Eventually, the communication probability among all CETIs could be estimated as 
\begin{equation}
    \mathcal{F}_{1,2} =  \frac{n_{1,2}}{C_{\rm N}^2}=\frac{2n_{1,2}}{N\times (N-1)},
\end{equation}
where $N$ is the number of CETIs and $n_{1}$ is the number of CETI pairs that can complete one-way communications and $n_{2}$ is the number of CETI pairs that can complete two-way communications. 

\section{The Simulation Results}
\label{3}
\subsection{The Possible Number of CETIs}
In our simulation, most initial setups are based on the latest astrophysical observations, invoking the star formation histories, metallicity distributions, and the likelihood of stars hosting
terrestrial planets in their habitable zones, etc. The most uncertain parameters are $f_c$ and $F$. Here we consider nine models, e.g. ($f_c=0.1\%,~F=25\%$), ($f_c=0.01\%,~F=25\%$), ($f_c=0.001\%,~F=25\%$), ($f_c=0.1\%,~F=45.7\%$), ($f_c=0.01\%,~F=45.7\%$), ($f_c=0.001\%,~F=45.7\%$), ($f_c=0.1\%,~F=75\%$), ($f_c=0.01\%,~F=75\%$), and  ($f_c=0.001\%,~F=75\%$). These models well include different situations that may occur, such as a high probability of terrestrial planets generating CETIs in the early/late stage of the host star's evolution history, or a low probability of terrestrial planets generating CETIs in the early/late stage of the host star's evolution history. 

\begin{figure*}[htb]
    \centering
  	\includegraphics[width=14cm,height=6cm]{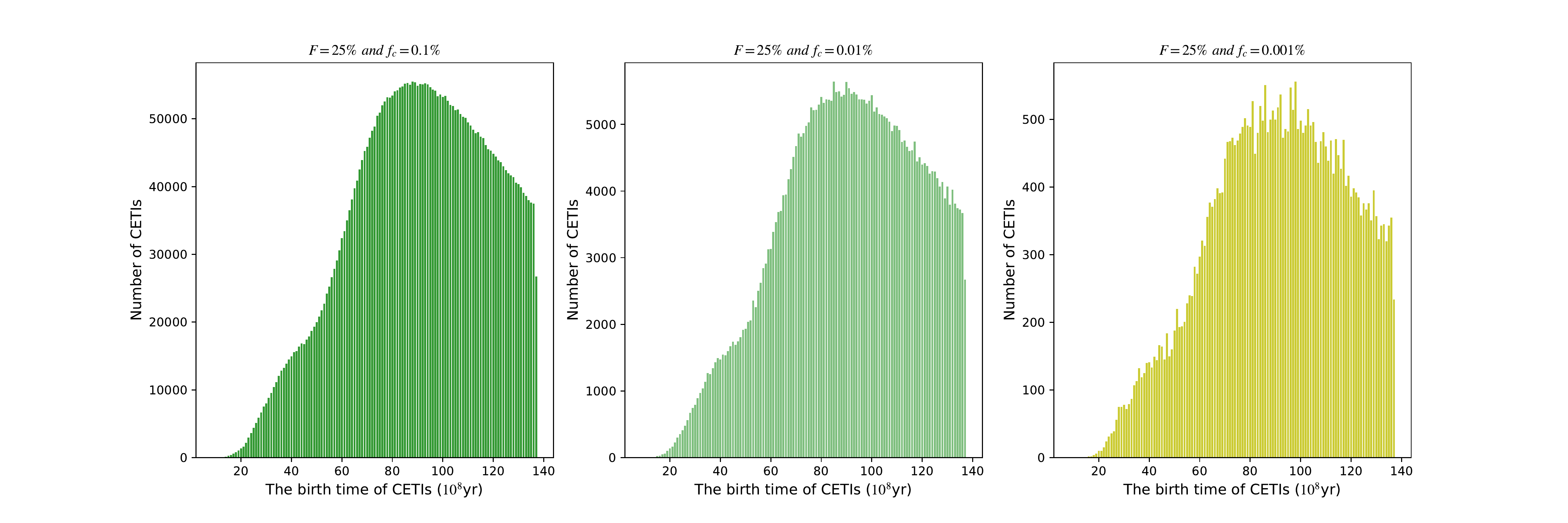} 
  	\includegraphics[width=14cm,height=6cm]{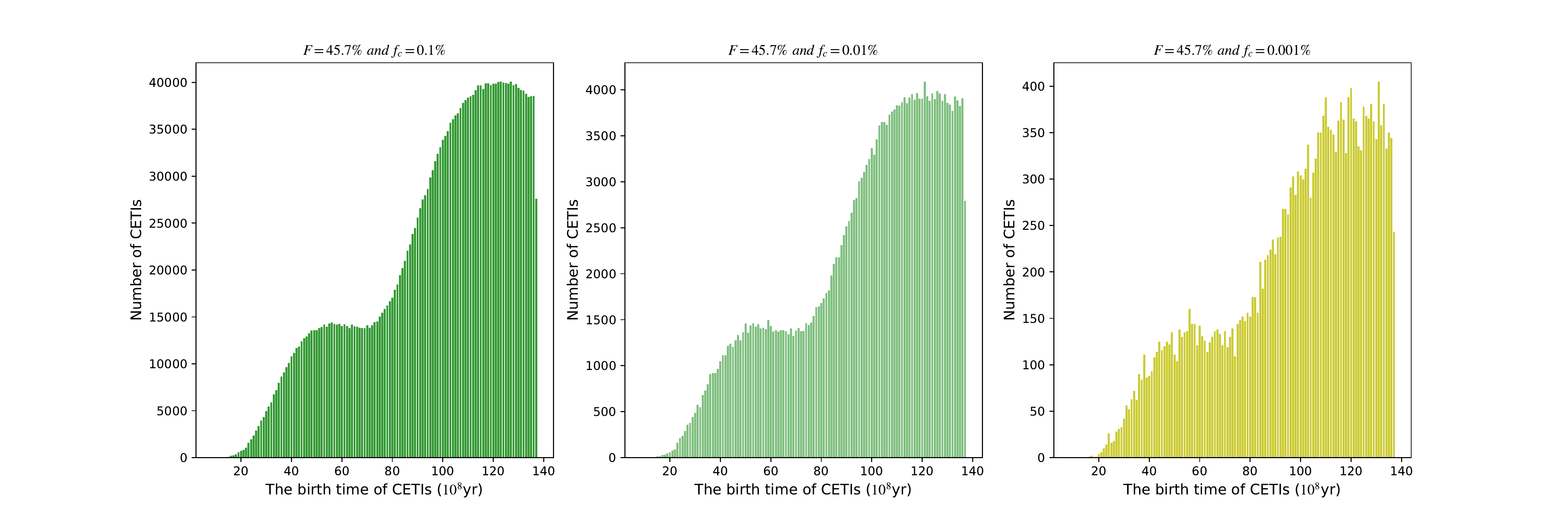} 
  	\includegraphics[width=14cm,height=6cm]{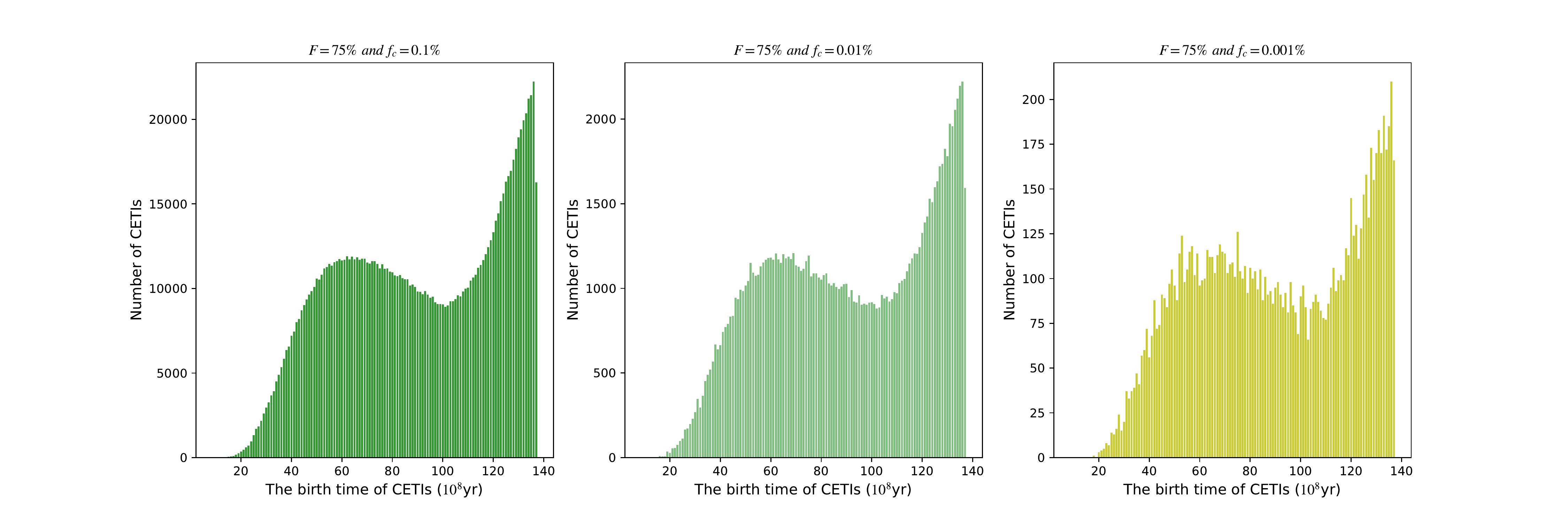} 
   \caption{Birth time distribution of CETIs for different models.}
   \label{birthtime}
\end{figure*}

For each model, we run the simulation 100 times and thus obtain an estimation for the possible number of CETIs under different models \footnote{The number of runs for each simulation is determined by balancing the computation time consumption and the resulting convergence.}. The results are collected in Table \ref{table 1}. We find that in an optimistic situation, the total number of CETIs that can be generated throughout the evolution history of the Milky Way could be in order of $10^4$, while under relatively conservative conditions, the total number of CETIs would be reduced to the order of 100. The value of $F$ would also affect the number of CETIs (the larger $F$ is, the fewer the CETIs). This is mainly because most stars in our simulation are low-mass M-type stars, which evolve slowly and have a longer lifetime ($\tau_i$). In this case, if these stars tend to generate CETIs in their late evolution stage (e.g. $F=75\%$), many CETIs have not yet been produced. In figure \ref{birthtime}, we plot the birth time distribution for all simulated CETIs. We show that when $F=25\%$, most CETIs would be born around 8.8 billion years after the birth of the universe. When $F=45.7\%$, most CETIs would be born around 12.3 billion years. Finally, when $F=75\%$, most CETIs would be born around 13.6 billion years.

\begin{table}[htb]
\caption{\rm The Number of CETIs That existed up till Now ($N$)}
	\centering
	\setlength{\tabcolsep}{4mm}
	\begin{tabular}{cccc}
	    \hline\hline\noalign{\smallskip}
		F & 25\% & 45.7\% & 75\% \\
		$f_c$ & & &\\
		\noalign{\smallskip}\hline\noalign{\smallskip}
		0.1\% & $42777_{-369}^{+267}$ & $26089_{-288}^{+226}$ & $12177_{-253}^{+283}$  \\
	     0.01\% & $4247_{-87}^{+72}$ & $2582_{-73}^{+79}$ & $1202_{-65}^{+70}$ \\
		 0.001\% & $393_{-20}^{+26}$ & $238_{-21}^{+33}$ & $111_{-17}^{+28}$ \\
		\noalign{\smallskip}\hline
	\end{tabular}
	\label{table 1}
\end{table}

\subsection{Communication Probability between CETIs}

\begin{figure*}[htb]
   \centering
   \includegraphics[width=14cm,height=7cm]{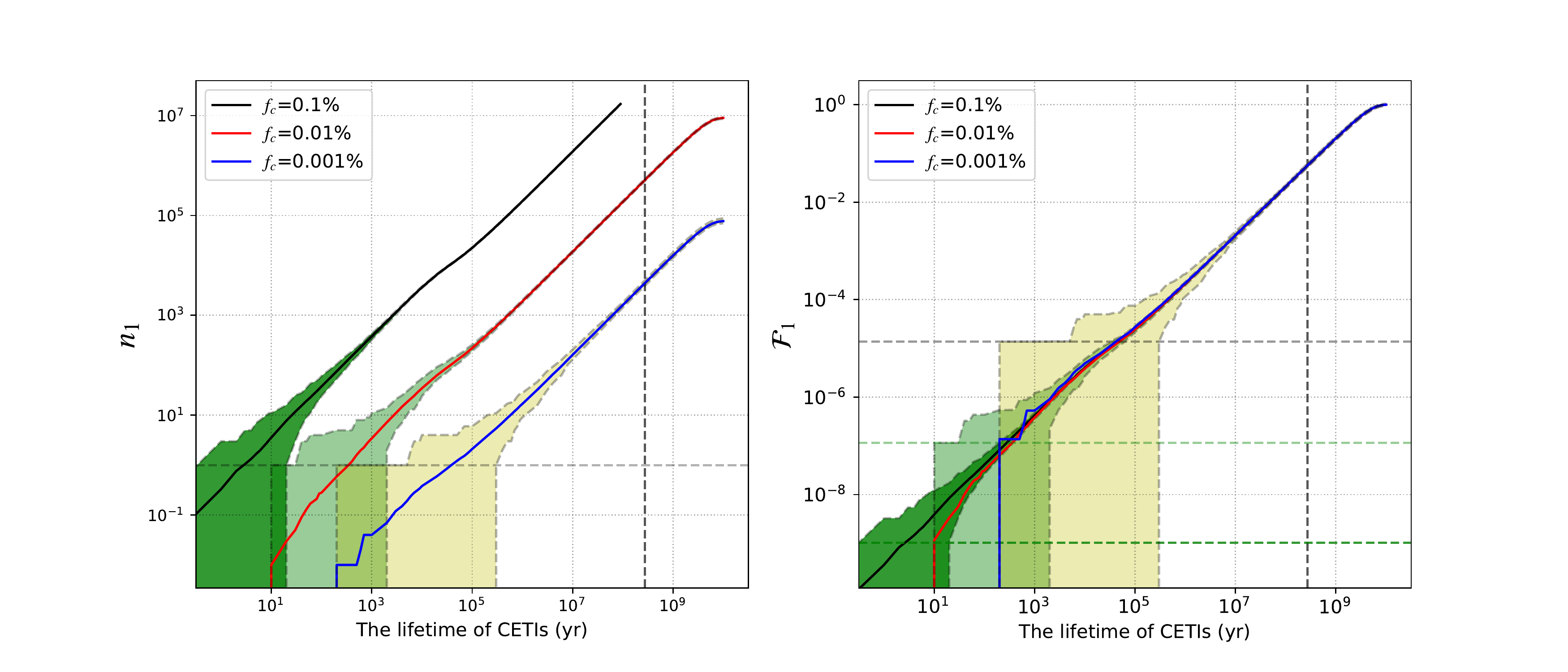} 
   \includegraphics[width=14cm,height=7cm]{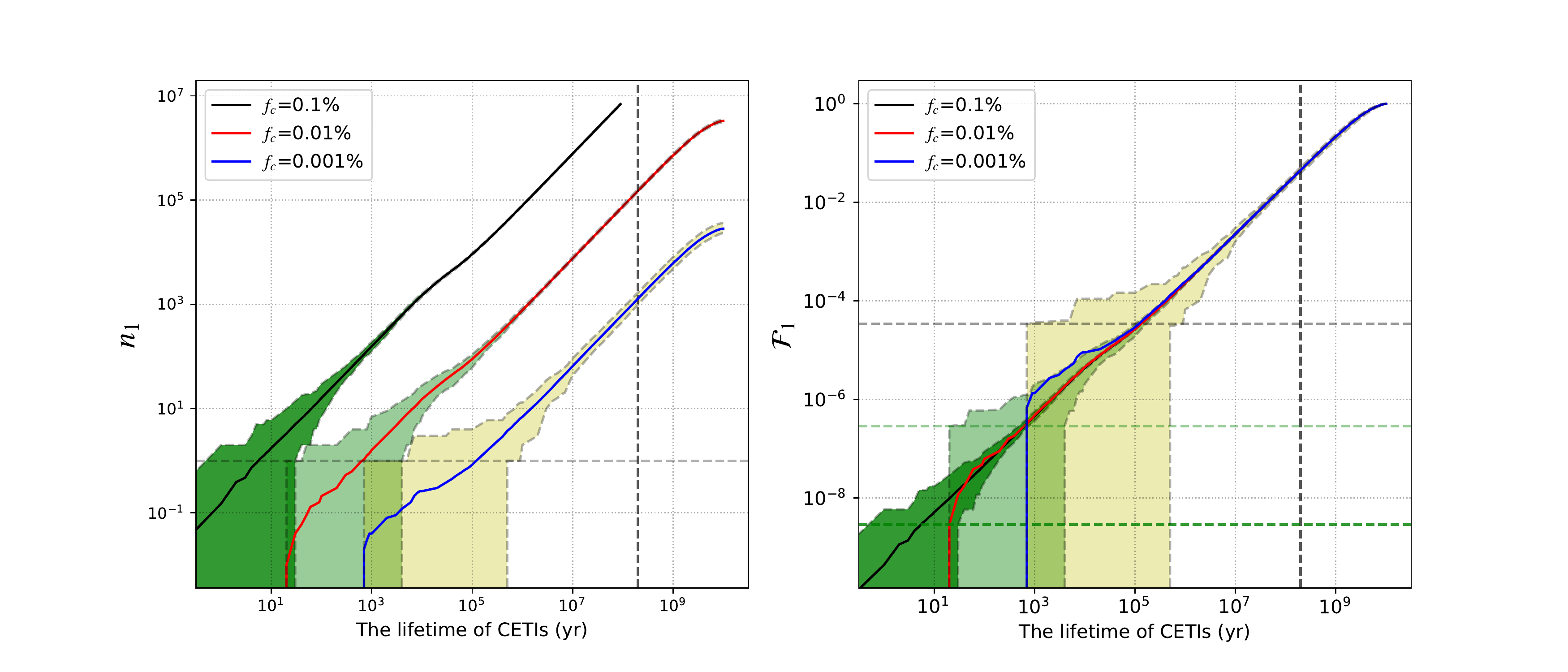} 
   \includegraphics[width=14cm,height=7cm]{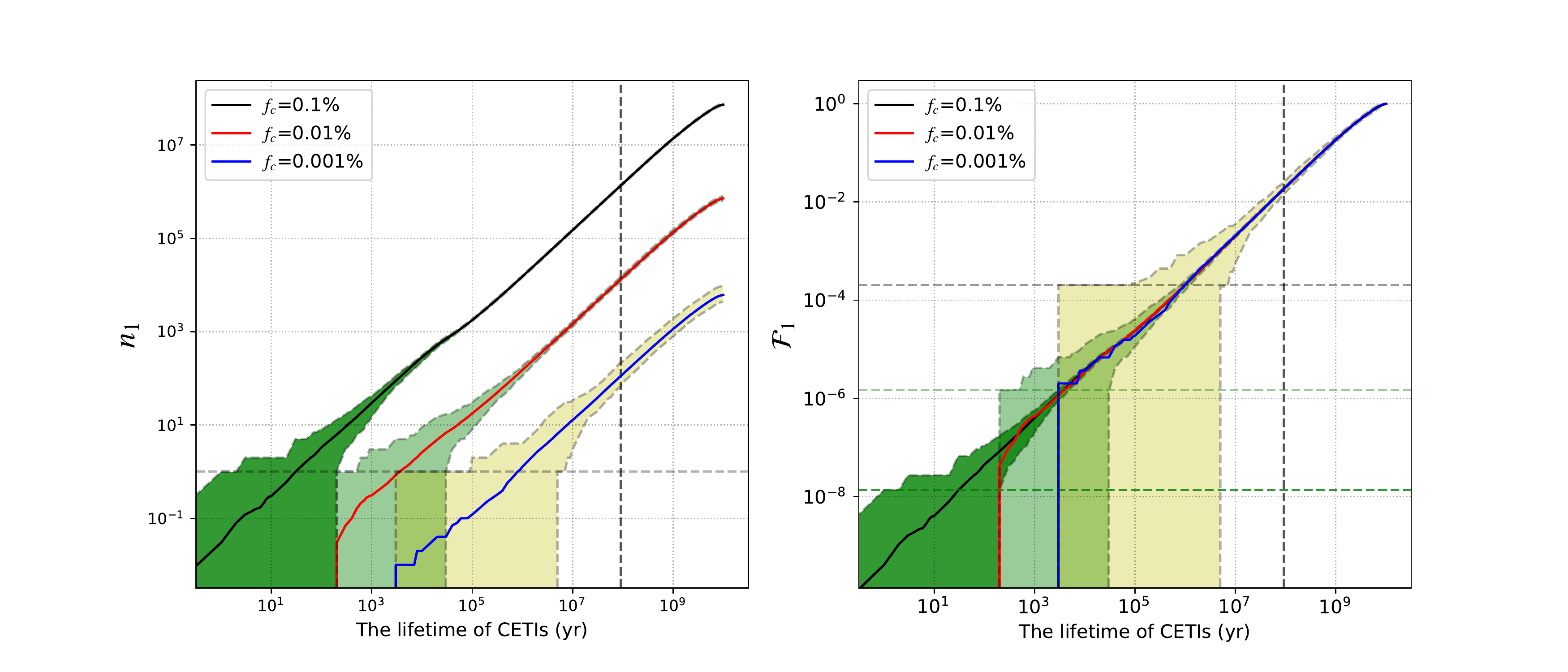} 
   \caption{Probability of one-way communication between CETIs. From top to bottom, panels correspond to the cases of $F=25\%$, $F=45.7\%$, and $F=75\%$, respectively. Left panels show how $n$ varies with $\Delta T$ and right panels show how $\mathcal{F}$ varies with $\Delta T$. The shaded area corresponds to the 1$\sigma$ error of the results. The dashed horizontal lines mark the result of $n_1=1$. The dotted vertical lines represent the upper limit of the lifetime of CETIs. }
   \label{oneway}
\end{figure*}

\begin{figure*}[htb]
   \centering
   \includegraphics[width=14cm,height=7cm]{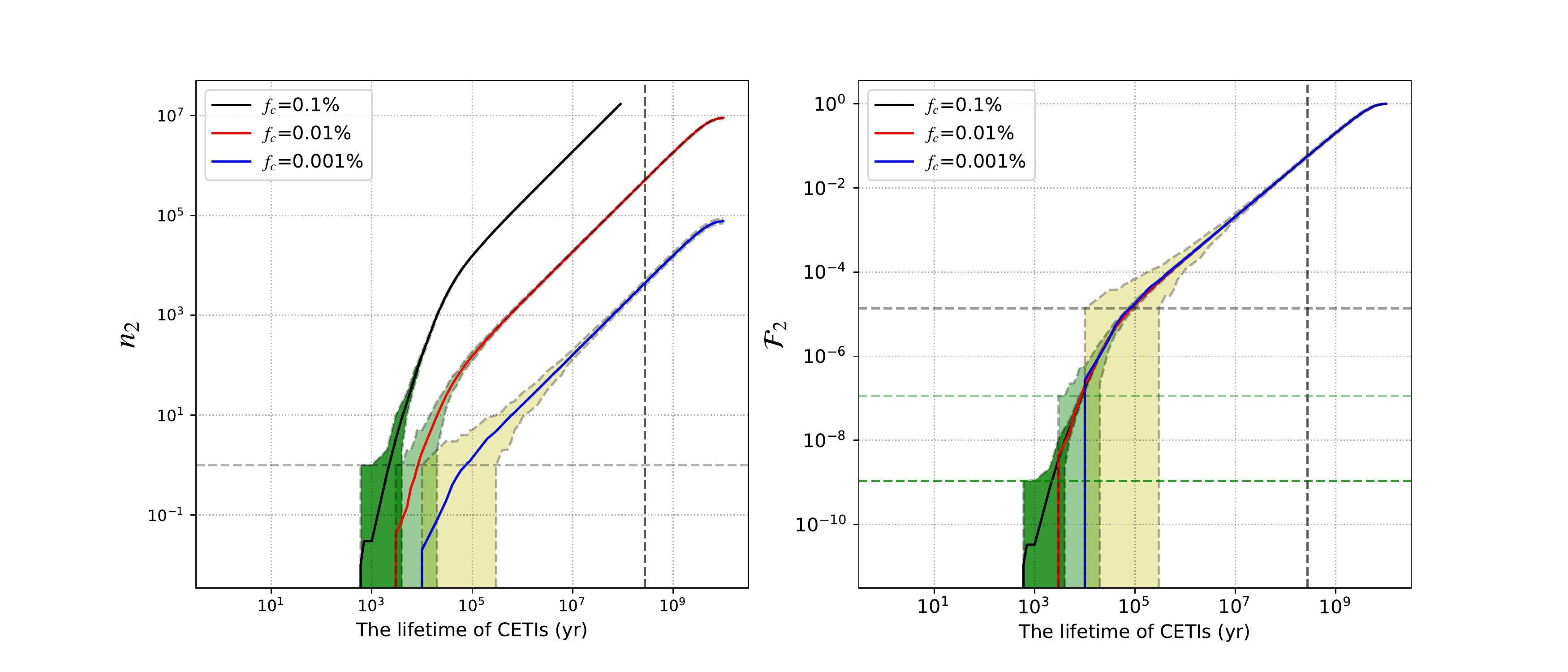} 
    \includegraphics[width=14cm,height=7cm]{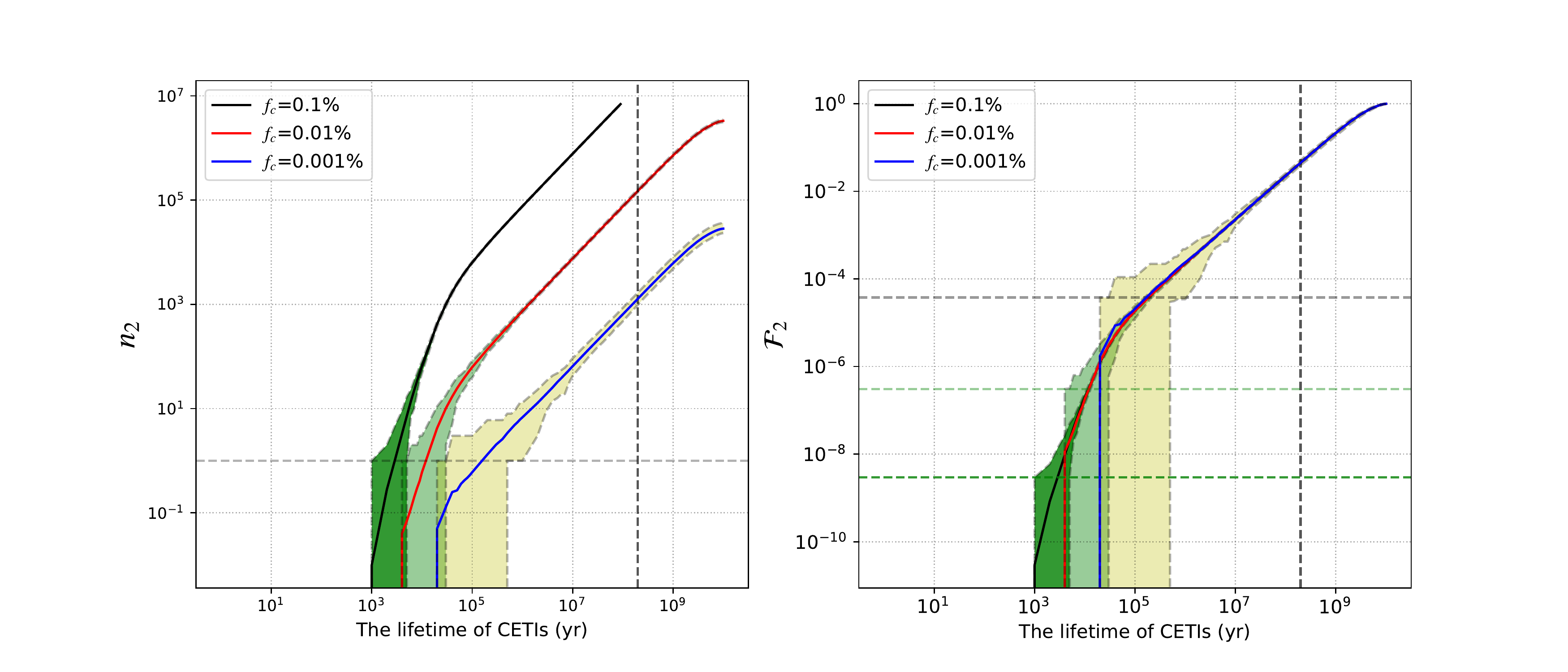} 
    \includegraphics[width=14cm,height=7cm]{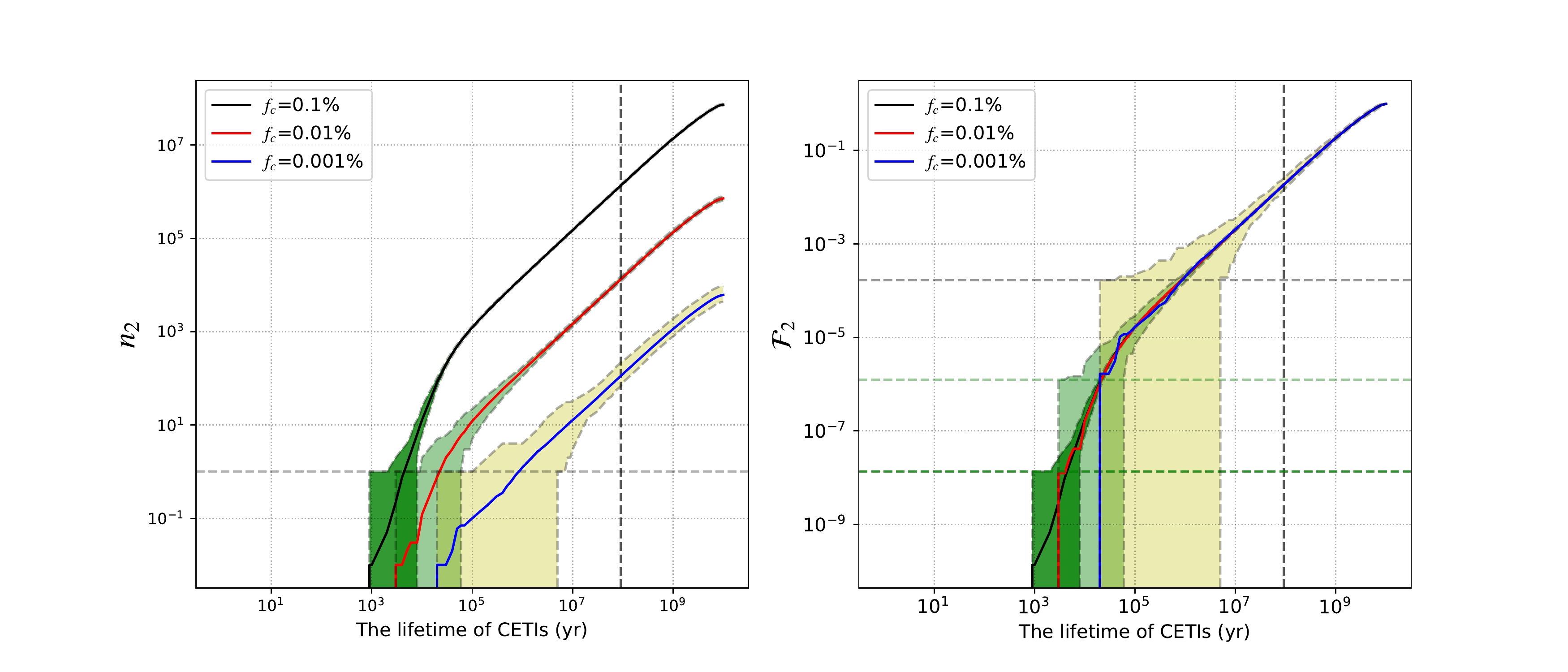} 
   \caption{Same as Figure \ref{oneway}, but for two-way communication between CETIs.}
   \label{twoway}
\end{figure*}

Here we choose a series of $\Delta T$ values for simulation, from 1yr to $10^9$ yr in log space. For each step, we run the simulation 100 times to make an estimation for the number of CETI pairs that can complete at least one time signal exchange ($n$) and the overall communication probability between CETIs ($\mathcal{F}$). In Figure \ref{birthtime}-\ref{earthoneway}, we plot the results that are relevant for reflecting the main conclusions. We find that both $n$ and $\cal{F}$ will increase with the increase of $\Delta T$. If the communication lifetime of CETIs is long enough, the communication probability between CETIs could approach unity. However, it is worth noting that there is a natural upper limit to the communication lifetime of each CETI, that is, $(1-F)\tau_i$, where $\tau_i$ is the main-sequence lifetime of its host star. For all simulated CETIs, we take the minimum value of $(1-F)\tau_i$ as the physical upper limit of $\Delta T$. For $F=25\%$, this upper limit would be around $2.7\times10^8$ yr, so the maximum value for $n_1$ and $\mathcal{F}_1$ would be $\sim10^7$/$5.6\times10^5$/$4.8\times10^3$ and $\sim10^{-2}$/$6.2\times10^{-2}$/$6.2\times10^{-2}$ (for $f_c=0.1\%/0.01\%/0.001\%$), respectively. For $F=45.7\%$, the upper limit of $\Delta T$ would be around $1.9\times10^8$ yr, so the maximum value for $n_1$ and $\mathcal{F}_1$ would be $\sim10^7$/$1.5\times10^5$/$1.3\times10^3$ and $\sim 10^{-2}$/$4.5\times10^{-2}$/$4.5\times10^{-5}$ (for $f_c=0.1\%/0.01\%/0.001\%$), respectively. For $F=75\%$, the upper limit of $\Delta T$ would be around $9.1\times10^7$ yr, so the maximum value for $n_1$ and $\mathcal{F}_1$ would be $1.4\times10^6$/$1.5\times10^4$/$2.2\times10^2$ and $1.8\times10^{-2}$/$1.8\times10^{-2}$/$1.8\times10^{-2}$ (for $f_c=0.1\%/0.01\%/0.001\%$), respectively. 

Considering many factors, such as the impact of the cosmic environment (e.g., the impact of high-energy events such as supernovae explosions), the impact of the planetary system environment (e.g., asteroid impacts), and the impact of the behavior of intelligent life itself (e.g., devastating civil war), the communication lifetime of CETIs should not be very long. In this case, one interesting question is what is the minimum value of $\Delta T$ to ensure that CETIs in the Milky Way could successfully communicate. Hereafter, we use $\Delta T_c$ to represent such a critical lifetime.

For one-way communication (i.e. $n_1=1$), the results of $\Delta T_c$ are summarized in Table \ref{table 2}. It is clear that with the decreasing of $f_c$ and the increasing of $F$, the critical lifetime $\Delta T_c$ will increase. Changing $F$ from $25\%$ to $75\%$ will result in a change of $\Delta T_c$ by several times to an order of magnitude. On the other hand, when $f_c$ decreases by an order of magnitude, it can result in $\Delta T_c$ increasing by two orders of magnitude or more. For the most optimistic situation (e.g. $F=25\%$ and $f_c=0.1\%$), the critical lifetime is only $3_{-2}^{+17}$ yr. If so, it is not difficult for CETIs to achieve one-way communication. However, for a pessimistic situation (e.g. $F=75\%$ and $f_c=0.001\%$), the critical lifetime needs to be $0.8_{-0.796}^{+1.2}$ Myr. If this is the case, it will be very difficult for CETIs to achieve one-way communication. 

As shown in figure \ref{twoway}, even in an optimistic situation (e.g. $f_c=0.1\%$), CETIs need to survive for thousands of years to achieve two-way communication. If $f_c=0.01\%$, CETIs need to survive for tens of thousands of years on average. More pessimistic is that when $f_c=0.001\%$, CETIs will need to survive for tens of thousands or even hundreds of thousands of years to achieve two-way communication.

\begin{table}[htb]
\caption{The Critical Value of Lifetime (yr) for $n_1=1$}
	\centering
	\setlength{\tabcolsep}{2mm}
	\begin{tabular}{cccc}
	    \hline\hline\noalign{\smallskip}
		F & 25\% & 45.7\% & 75\% \\
		$f_c$ & & &\\
		\noalign{\smallskip}\hline\noalign{\smallskip}
		0.1\% & $3_{-2}^{+17}$ & $6_{-5}^{+24}$ & $40_{-39}^{+160}$  \\
	    0.01\% & $400_{-390}^{+1600}$ & $700_{-680}^{+3300}$ & $4000_{-3800}^{+26000}$ \\
		0.001\% & $40000_{-38000}^{+260000}$ & $200000_{-199300}^{+300000}$ & $800000_{-796000}^{+1200000}$ \\
	    \noalign{\smallskip}\hline
	\end{tabular}
	\label{table 2}
\end{table}


\subsection{The Possibility of Humans Receiving Signals from Other CETIs}

\begin{figure*}[htb]
   \centering
   \includegraphics[width=14cm,height=7cm]{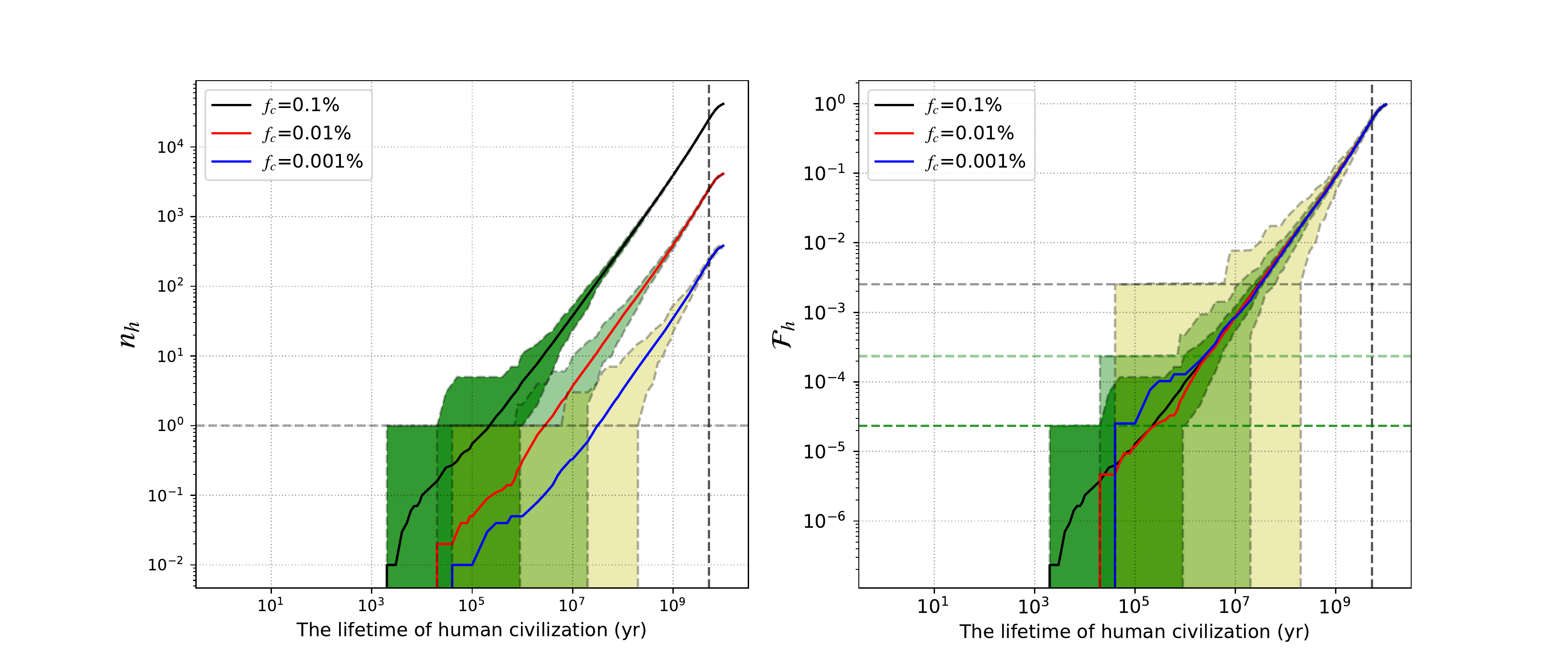} 
    \includegraphics[width=14cm,height=7cm]{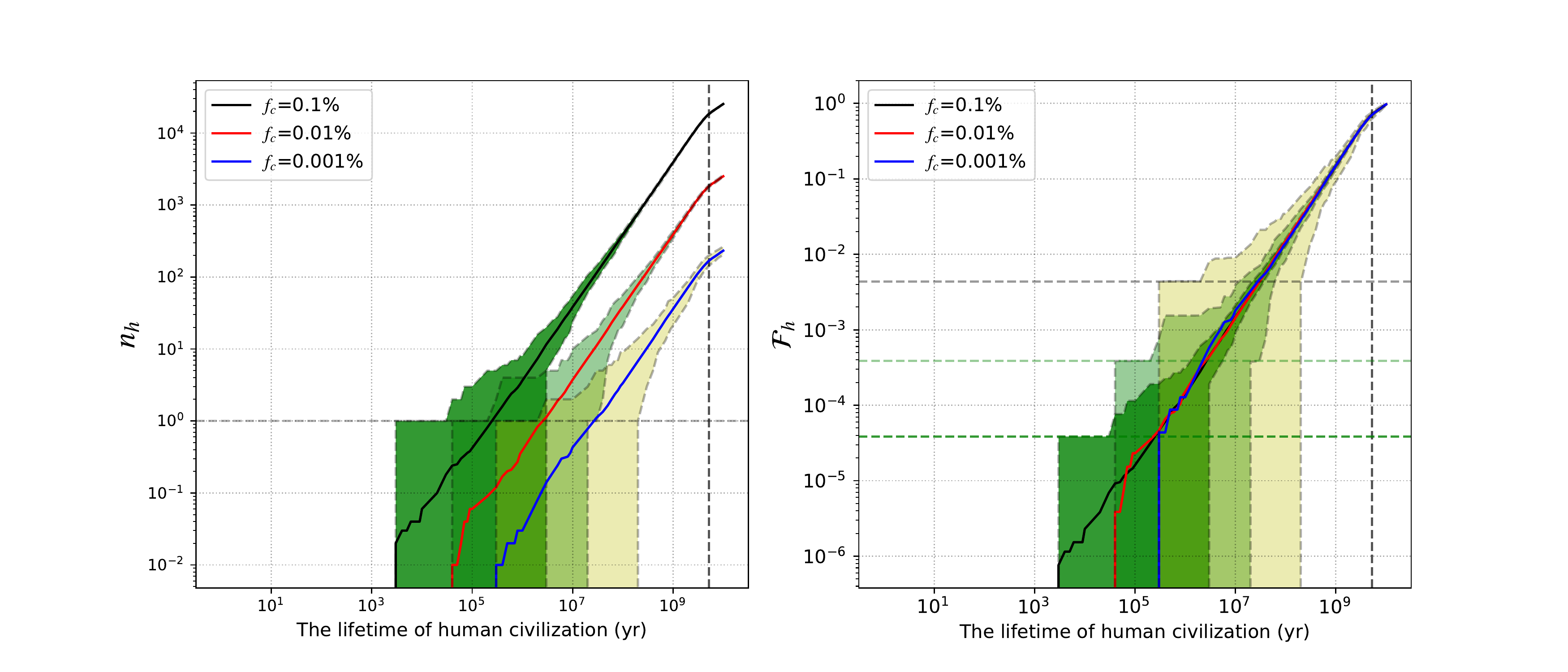} 
    \includegraphics[width=14cm,height=7cm]{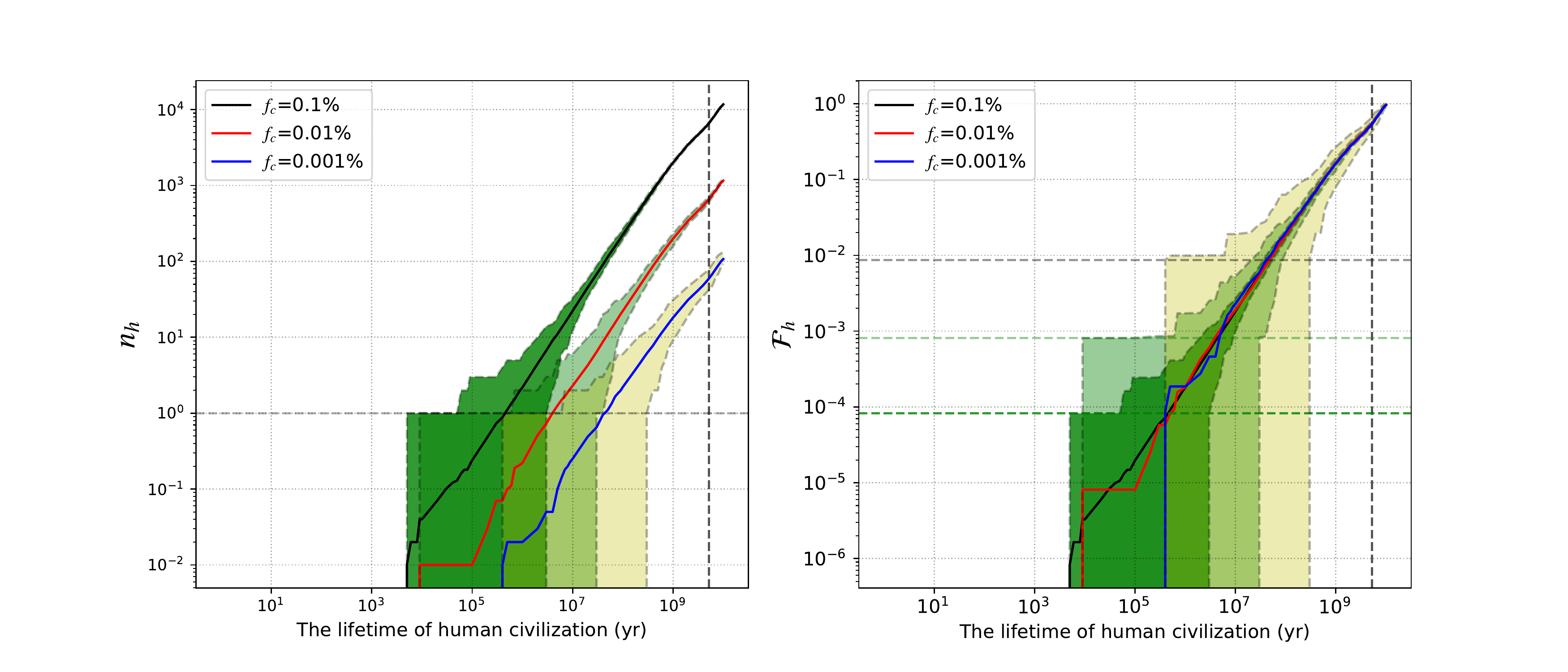} 
   \caption{Same as Figure \ref{oneway}, but for the probability of humans receiving alien signals. 
   \label{earthoneway}
   }
\end{figure*}

As the only known CETI, the question we are most interested in is how long we need to survive in order to receive signals from other CETIs. With the help of our previous simulation results, we have carried out a detailed study on this problem. First, for each simulation run, we manually add one more data point, ($X_{\rm h},Y_{\rm h},Z_{\rm h},T_{\rm h}$), where ($X_{\rm h},Y_{\rm h},Z_{\rm h}$) are the positions of the sun in the Milky Way, and $T_{\rm h}=137.72$ is the time when humans become a CETI (roughly today). We then calculate the probability of humans receiving signals as 
\begin{equation}
    \mathcal{F}_{\rm h} = \frac{n_{\rm h}}{N},
\end{equation}
where $N$ is the total number of CETIs, and $n_{\rm h}$ is the number of CETIs whose signal could be received by humans. In Figure \ref{earthoneway}, we show how $n_{\rm h}$ and $\mathcal{F}_{\rm h}$ vary with respect to $\Delta T$. Here we take $(1-F)\tau_\odot$ as the physical upper limit of $\Delta T$, where $\tau_\odot =10 Gyr$ is the main-sequence lifetime of the sun. For $F=25\%$, this upper limit would be around $5.4\times10^9$ yr, so the maximum value for $n_{\rm h}$ and $\mathcal{F}_{\rm h}$ would be $2.4\times10^4$/$2.4\times10^3$/$2.2\times10^2$ and $5.6\times10^{-1}$/$5.6\times10^{-1}$/$5.6\times10^{-1}$ (for $f_c=0.1\%/0.01\%/0.001\%$), respectively. For $F=45.7\%$, the upper limit of $\Delta T$ would be around $5.4\times10^9$ yrs, so the maximum value for $n_{\rm h}$ and $\mathcal{F}_{\rm h}$ would be $1.8\times10^4$/$1.8\times10^3$/$1.6\times10^2$ and $6.9\times10^{-1}$/$6.9\times10^{-1}$/$6.9\times10^{-1}$ (for $f_c=0.1\%/0.01\%/0.001\%$), respectively. For $F=75\%$, the upper limit of $\Delta T$ would be around $5.2\times10^9$ yr, so the maximum value for $n_{\rm h}$ and $\mathcal{F}_{\rm h}$ would be $6.4\times10^3$/$6.4\times10^2$/$5.7\times10^1$ and $5.3\times10^{-1}$/$5.3\times10^{-1}$/$5.2\times10^{-1}$ (for $f_c=0.1\%/0.01\%/0.001\%$), respectively.

We also calculate the critical lifetime to ensure that humans could successfully receive at least one signal. The results for different cases are collected in Table \ref{table 3}. We find that the change of $F$ value would barely affect the results, while when $f_c$ decreases by an order of magnitude, it can result in $\Delta T_c$ increasing by one order of magnitude. For the most optimistic situation (e.g. $F=25\%$ and $f_c=0.1\%$), the critical lifetime is about $0.3_{-0.298}^{+0.6}$ Myr. Considering the randomness of the location and birth time of CETIs, we can take the minimum $\Delta T_c$ in the range of 1$\sigma$ as the most optimistic estimate; even so, humans need to stay as a CETI for at least 2000 yr to receive one signal. For pessimistic situations (e.g. $F=75\%$ and $f_c=0.001\%$), the critical lifetime is about $50_{-49.6}^{+250}$ Myr and, even with the minimum $\Delta T_c$ in the range of 1$\sigma$, could reach 400,000 yr. In this case, it will be extremely difficult for humans to communication with other CETIs.

\begin{table}[htb]
\caption{The Critical Value of Lifetime (Myr) for $n_{\rm{h}}=1$}
	\centering
	\setlength{\tabcolsep}{4.5mm}
	\begin{tabular}{cccc}
	    \hline\hline\noalign{\smallskip}
		F & 25\% & 45.7\% & 75\% \\
		$f_c$ & & &\\
		\noalign{\smallskip}\hline\noalign{\smallskip}
		0.1\% & $0.3_{-0.298}^{+0.6}$ & $0.3_{-0.297}^{+1.7}$ & $0.5_{-0.495}^{+2.5}$  \\
		0.01\% & $3_{-2.98}^{+17}$ & $3_{-2.96}^{+20}$ & $4_{-3.992}^{+26}$ \\
		0.001\% & $40_{-39.96}^{+160}$ & $30_{-29.70}^{+200}$ & $50_{-49.6}^{+250}$ \\
		\noalign{\smallskip}\hline
	\end{tabular}
	\label{table 3}
\end{table}

\section{Conclusions and Discussion}
\label{4}
In this paper, we carry out Monte Carlo simulations to calculate the number of possible communicating extraterrestrial intelligent civilizations within our Galaxy and the communication probability among these CETIs based on the latest astrophysical information. We also calculate the likelihood that humans would receive signals from other CETIs. We find that for optimistic situations (e.g. $F=25\%$ and $f_c=0.1\%$), there could be $42777_{-369}^{+267}$ CETIs and they need to survive for $3_{-2}^{+17}$ yr ($2000_{-1400}^{+2000}$ yr) to achieve one way communication (two way communication). However, for pessimistic situations (e.g. $F=75\%$ and $f_c=0.001\%$), only $111_{-17}^{+28}$ CETIs could be born and they need to survive for $0.8_{-0.796}^{+1.2}$ Myr ($0.9_{-0.88}^{+4.1}$ Myr) to achieve one-way communication (two-way communication). Our simulation may also provide a quantitative explanation for the so-called Fermi Paradox, the question was about why we have not received signals from CETIs. The results indicate even if there are $42777_{-369}^{+267}$ CETIs ever existing in the Milky Way, we still need to survive at least 2000 yr to receive a signal. While if there are only $111_{-17}^{+28}$ CETIs ever existing, we still need to survive at least 400,000 yr to receive a signal. The reason why we have not received a signal may be that the communication lifetime of human is not long enough at present. However, it has been proposed that the lifetime of civilizations is very likely self-limiting (known as the Doomsday argument), due to many potential disruptions, such as population issues, nuclear annihilation, sudden climate change, rouge comets, ecological changes, etc \citep{2019Bloetscher}. If the Doomsday argument is correct, for some pessimistic situations, humans may not receive any signals from other CETIs before extinction. Nevertheless, it is also possible that no two CETIs in the Milky Way could ever communicate with each other. However, it is worth noticing that due to the large uncertainty of the variables in the Drake equation, the tails in the simulated distribution may provide more interest than the average \citep{2019Bloetscher}. According to our simulations, for the tail value of some optimistic situations, human beings still have hope of detecting a CETI signal.

Here we would like to discuss some caveats about our numerical method and how these might affect the results. First of all, the values of $f_c$ and $F$ are full of many unknowns. It is quite uncertain what proportion of terrestrial planets can give birth to life, and the process of life evolving into a CETI and being able to send detectable signals to space is highly unpredictable. Some devastating disasters may prevent life from evolving into a CETI or even annihilate them at an early stage. For example, the dinosaurs that ruled the Earth for more than 100 million years died out when an asteroid hit the Earth 65 million years ago. So far, we do not have enough observational data to provide a precise estimation for $f_c$. The minimum value ($0.001\%$) we take may also be overestimated. If so, the number of CETIs would become even lower, and the opportunities for communication between CETIs would become extremely small. On the other hand, here we assume that the appearance time of CETIs is proportional to the lifetime of their host stars by introducing the $F$ parameter. This hypothesis is formed based on the consideration that the lifetime of stars is related to their mass. The life of a massive star is short, but its violent evolution will make the surrounding space environment more complex, which is more conducive to accelerating the evolution of complex systems and therefore, CETIs may appear earlier. Although we are completely unclear about the specific value of $F$, we have considered three cases of $F$ from large to small, which should be able to cover the real situation. Another view is that the occurrence time of CETIs should be an absolute value independent of the lifetime of their host stars. Based on the only known data point (human beings), the absolute value is usually taken as $\sim 5~\rm Gyr$. If this is the true case, considering the average lifetime of FGKM stars, our results with $F=25\%$ (or maybe with an even smaller $F$) should be relevant. Note that based on such assumptions, some previous works \citep[e.g.][]{2019Bloetscher,2020ApJ...896...58W,zhang2020quantitative} have made a good estimate for the current number of CETIs within our Galaxy. Their results are comparable with our ($f_c=0.001\%,~F=25\%$) model in orders of magnitude\footnote{The number of CETIs in our results is higher, because we focus on the number of CETIs that have ever existed up until now, but previous works focus on the number of CETIs that currently exist.}.

\par Secondly, we did not consider the details of signal transmission and reception here. Most of the current research is biased toward assuming that CETIs may have the ability to transmit radio signals like human beings. However, the propagation of radio in cosmic space will cause a decrease of flux as a function of distance. Meanwhile, the signal may be reflected or refracted when passing through the medium.
For example, in 1974, in order to celebrate the establishment of the Arecibo telescope, humans transmitted a string of 1679 binary digits to the globular cluster M13, called the “Arecibo message”. The power of this signal is 450 kW and the duration is 169 s. However, at the distance of M13($\sim$ 25,000 lt-yr away), the flux density of the “Arecibo message” is $\sim 2.7 \times 10^{-20} \rm Jy$, which is 21 orders of magnitude fainter than the sensitivity (10 Jy) of the Allen Telescope Array (the narrowband SETI search instrument currently used by humans) \citep{zhang2020quantitative}. If there are indeed CETIs in M13, their detection ability needs to be 21 orders of magnitude higher than ours to detect our signal. Conversely, if they transmit a similar signal, we need to improve the detection ability by 21 orders of magnitude to detect it. The problem of signal attenuation will further reduce the probability of communication between CETIs. Recently, \citet{zhang2020quantitative} proposed that a new view that CETIs may send FRB-like artificial signals to the Galaxy for communication. If so, the probability of communication between CETIs might be enhanced (see details in their paper). 
\par Thirdly, here we assume that the signals emitted by all CETIs were isotropic and assume that CETIs have been transmitting signals all the time during their communication lifetime. In reality, CETIs may transmit signals intermittently, and may have certain directional selectivity in order to improve their signal power. If so, the probability of communication between CETIs would be reduced.
\par Finally, in the literature, the definition for habitable zones is still debated. Here we assume that the existence of liquid water on the surface of the planet is a necessary condition for a habitable zone. However, some works suggest that other CETIs may not require water but need other solvents to facilitate chemical reactions\citep{2018arXiv180207036G}. On the other hand, besides the concept of a “habitable zone” for planets, some works propose the concept of a galactic habitable zone, which is an annular region between 7 and 9 kpc from the center of the Milky Way and widens over time \citep{Lineweaver2004The}. The uncertainty over “habitable zones” will increase the uncertainty of estimations for the number of CETIs and the communication probability between CETIs. 

At the end of this paper, we want to emphasize that with the rapid progress of astronomical observation, the quantitative calculation in the study of CETIs has become possible. Although there are still some uncertainties in the calculation results, these can be overcome with the further improvement of astronomical observation ability in the future. Only with these quantitative calculation results can we effectively judge whether some suspected signals are indeed sent by other CETIs, and use the detected signals to develop valuable research on other CETIs, such as the birth rate and lifetime of CETIs ($f_c$, $\Delta T$), and so on. 
\acknowledgments
We thank the anonymous referee for the helpful comments that have helped us to improve the presentation of this paper. This work is supported by the National Natural Science Foundation of China (NSFC) under grant No. 12021003.

\bibliography{reference}
\begin{appendix}
In order to facilitate readers' understanding of each parameter, we summarize the parameters and their definitions mentioned in this article in Table \ref{table 4}.
\begin{table}[htbp]
\caption{Notation List for Different Parameters Shown in This Work.}
	\renewcommand\arraystretch{1.3}
	\centering
	\scriptsize
	\setlength{\tabcolsep}{4mm}
	\begin{tabular}{|l|c|}
	    \hline Symbol & Definition\\
		\hline $f_p$ & the average probability of terrestrial planets appearing in the habitable zone of stars.\\
		\hline $f_c$ & the probability of life appearing on terrestrial planets and eventually evolving into a CETI. \\
		\hline $T_s$ & the birth time of stars.\\
		\hline $T_c$ & the birth time of CETIs.\\
		\hline $F$ & the parameter that decides at what stage of star evolution a CETI was born.\\
		\hline $\tau_i$ & the main-sequence lifetime of stars.\\
		\hline $\Delta t$ & the simulation time-step, $10^8$ yr.\\
		\hline $\Delta T$ & the communication lifetime of CETIs.\\
		\hline $\Delta T_{\rm c}$ & the minimum value of $\Delta T$ to ensure that CETIs in the Milky Way could successfully communicate.\\
		\hline $N$ & the total number of CETIs.\\
		\hline $D_{\rm dis}$ & the distance between CETIs.\\
		\hline $T_{\rm sp}$ & the spread time of a signal.\\
		\hline $n_{1}$ & the number of CETI pairs that can complete one-way communications.\\
		\hline $n_{2}$ & the number of CETI pairs that can complete two-way communications.\\
		\hline $n_{\rm h}$ & the number of CETIs whose signal could be received by humans.\\
		\hline $\mathcal{F}_{1}$ & the probability of one-way communication. \\
		\hline $\mathcal{F}_{2}$ & the probability of two-way communication. \\
		\hline $\mathcal{F}_{\rm h}$ & the possibility of humans receiving signals from other CETIs.\\
		\hline
	\end{tabular}
	\label{table 4}
\end{table}
\end{appendix}
\end{document}